\documentclass{iau}
\usepackage{graphicx}

\title[Giant planet formation: planetesimal fragmentation and planet migration] 
{Planetesimal fragmentation and giant planet formation: the role of planet migration}

\author[Guilera, Swoboda, Alibert, de El\'{\i}a, Santamar\'{\i}a \& Brunini] 
{O. M. Guilera$^{1}$, D. Swoboda$^{2}$, Y. Alibert$^{2,3}$, G. C. de El\'{\i}a$^{1}$, P. J. Santamar\'{\i}a$^{1}$ \& A. Brunini$^{1}$}

\affiliation{$^1$Grupo de Ciencias Planetarias, Facultad de Ciencias Astron\'omicas y Geof\'{\i}sicas \& Instituto de Astrof\'{\i}sica de La Plata (Consejo Nacional de Investigaciones Cient\'{\i}ficas y T\'ecnicas - Universidad Nacional de La Plata), Argentina. \\[\affilskip]
$^2$ Physics Institute and Center for Space and Habitability, University of Bern, Bern, Switzerland. \\[\affilskip]  
$^3$ Observatoire de Besan{\c c}on, France. \\email: {\tt oguilera@fcaglp.unlp.edu.ar} }

\pubyear{2014}
\volume{310}  
\pagerange{xxx--xxx}
\jname{Complex Planetary Systems}
\editors{Z. Knezevic \& A. Lema\^itre}
\begin{document}

\maketitle

\begin{abstract}
In the standard model of core accretion, the cores of the giant planets form by the accretion of planetesimals. In this scenario, the evolution of the planetesimal population plays an important role in the formation of massive cores. Recently, we studied the role of planetesimal fragmentation in the in situ formation of a giant planet. However, the exchange of angular momentum between the planet and the gaseous disk causes the migration of the planet in the disk. In this new work, we incorporate the migration of the planet and globally study the role of planet migration in the formation of a massive core when the population of planetesimals evolves by planet accretion, migration due to the nebular drag, and fragmentation due to planetesimal collisions.        
 
\keywords{giant planet formation, planetesimal fragmentation, planet migration}
\end{abstract}

\firstsection 

 \vspace*{-.4 cm}

\section{Introduction}

In the standard model of core accretion, the formation of a giant planet ocurrs by four principal stages (\cite[Pollack et al. 1996]{Pollack.et.al.1996}, \cite[Fortier et al. 2009]{Fortier.et.al.2009}): 
\begin{itemize}
\item[i-] first, a solid core is formed by the accretion of planetesimals,
\item[ii-] this solid core binds the surrounding gas and a gaseous envelope grows in hydrostatic equilibrium, 
\item[iii-] initially, the planetesimal accretion rate is higher than the gas accretion rate, so the solid core grows faster than the gaseous envelope, but when the mass of the envelope equals the mass of the core\footnote{It is often said that the mass of the core reaches a critical value} the planet triggers the gas accretion and big quantities of gas are accreted in a short period of time, 
\item[iv-] finally, for some mechanism poorly understood the planet stops the accretion of gas and evolves in isolation, contracting and cooling at constant mass.   
\end{itemize}

The mass of the core to start the gaseous runaway phase is found to be
$\gtrsim 10~\mathrm{M}_{\oplus}$ (although, recent works showed that if
the envelope's grain opacity is lower than the values of the ISM
(\cite[Movshovitz et al. 2010]{[Movshovitz.et.al.2010}) or if there is
an increment of the envelope's abundance of heavy elements (\cite[Hori
\& Ikoma, 2011]{Hori.and.Ikoma.2011}), the critical core mass could be much lower than in the classical scenario). So, the real bottleneck for giant planet formation in the core accretion model, is the growth of the critical core mass before the dissipation of the disk. In a recent work (\cite[Guilera et al. 2014]{Guilera.et.al.2014}), we found that if planetesimal fragmentation is taken into account, the formation of massive cores in a few millon years is only possible starting with a population of big planetesimals (of 100~km of radius) and massive disks, and if most of the mass loss in planetesimal collisions is distributed in larger fragments. However, in this work the migration of the planet is neglected. The exchange of angular momentum between the planet and the gaseous disk forces the planet to migrate along the disk, entering in new zones of the population of planetesimals which could help in the formation of a massive core (\cite[Alibert et al. 2005 b]{Alibert.et.al.2005b}). In this new work, we incorporate type I migration in our global model of giant planet formation (\cite[Guilera et al. 2010]{Guilera.et.al.2010}).  

 \vspace*{-.4 cm}

\section{The model}

Following the work of \cite[Alibert et al. (2005 a)]{Alibert.et.al.2005a}, we incorporated type I migration in our model of giant planet formation. We also used the prescription derived by \cite[Tanaka et al. (2002)]{Tanaka.et.al.2002}, with an ad-hoc reduction factor, to calculate the velocity migration of the planet given by:
\begin{eqnarray}
  \frac{da_P}{dt}= -2~f_I~a_P\frac{\Gamma}{\textrm{L}_P},
\end{eqnarray}
where $a_P$ represents the planet's semi-major axis, $f_I$ is the reduction factor, and $\textrm{L}_P= \textrm{M}_P \sqrt{G \textrm{M}_{\star} a_P}$ is the angular momentum of the planet. $\Gamma$ is the total torque, which is given by:
\begin{eqnarray}
  \Gamma= (1.364 + 0.541 \alpha) \left( \frac{\textrm{M}_P ~ a_P ~ \Omega_P}{\textrm{M}_{\star} c_{s_P}} \right)^2 \Sigma_P ~ a_P^4 ~ \Omega_P^2,
\end{eqnarray}
where $\Omega_P$, $ c_{s_P}$ and $\Sigma_P$ are the values of the
keplerian frequency, the sound speed, and the gas surface
density at the position of the planet, respectively. The factor $\alpha$
is defined by:
\begin{eqnarray}
 \alpha= \frac{d\log\Sigma}{d\log R} \bigg|_{R= a_P}, \mbox{$R$ being the radial coordinate.}
\end{eqnarray}
The rest of the model is the same as the one described in Guilera et al. (2010,
2011, 2014).

\vspace*{-.4 cm}

\section{Results}

We studied the formation of a giant planet (until the planet's core
reaches the critical mass) with an initial semi-major axis of 5~au. We
focused on the comparison of two cases: the in situ formation of
the planet, and when the planet migrates in the disk under type I
migration. We considered an initial homogeneous population of
planetesimals of 100~km of radius and a disk ten times more massive than
the Minimum Mass Solar Nebula (\cite[Hayashi, 1981]{Hayashi.1981}). As in
\cite[Guilera et al. (2014)]{Guilera.et.al.2014}, we carried out two sets of simulations: when the population of planetesimals evolves by planet
accretion and planetesimal migration (hereafter case a), and when the population of planetesimals evolves by planet accretion, planetesimal migration and
planetesimal fragmentation (hereafter case b).

\begin{figure}[t]
\begin{center}
 \includegraphics[scale= 0.565]{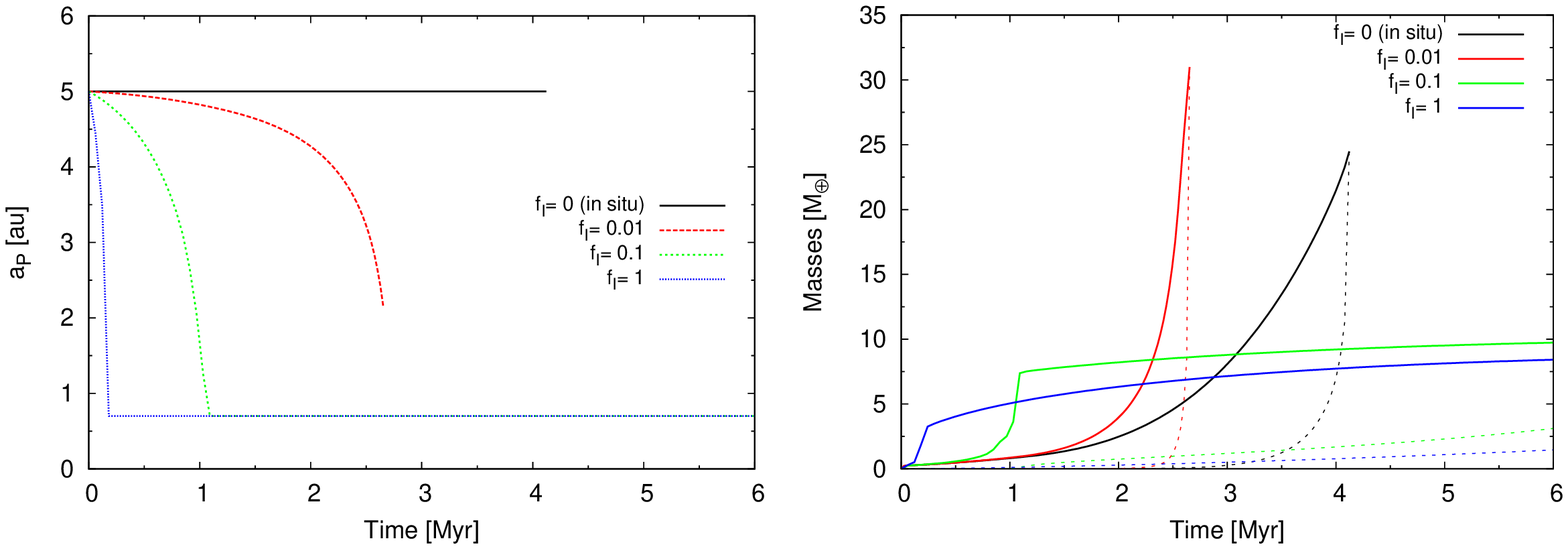} 
 \vspace*{-.6 cm}
 \caption{Left panel: time evolution of the planet's semi-mayor axis. Right panel: time evolution of the core mass (solid line) and envelope mass (dashed line). Both cases correspond to an initial embryo of $0.005~\textrm{M}_{\oplus}$ located at 5 au, and for diferent values of the ad hoc reduction factor of the planet's velocity migration. Color figure only available in the electronic version.}
 \label{fig1}
\end{center}
\end{figure}

\begin{figure}[t]
\begin{center}
 \includegraphics[scale= 0.535]{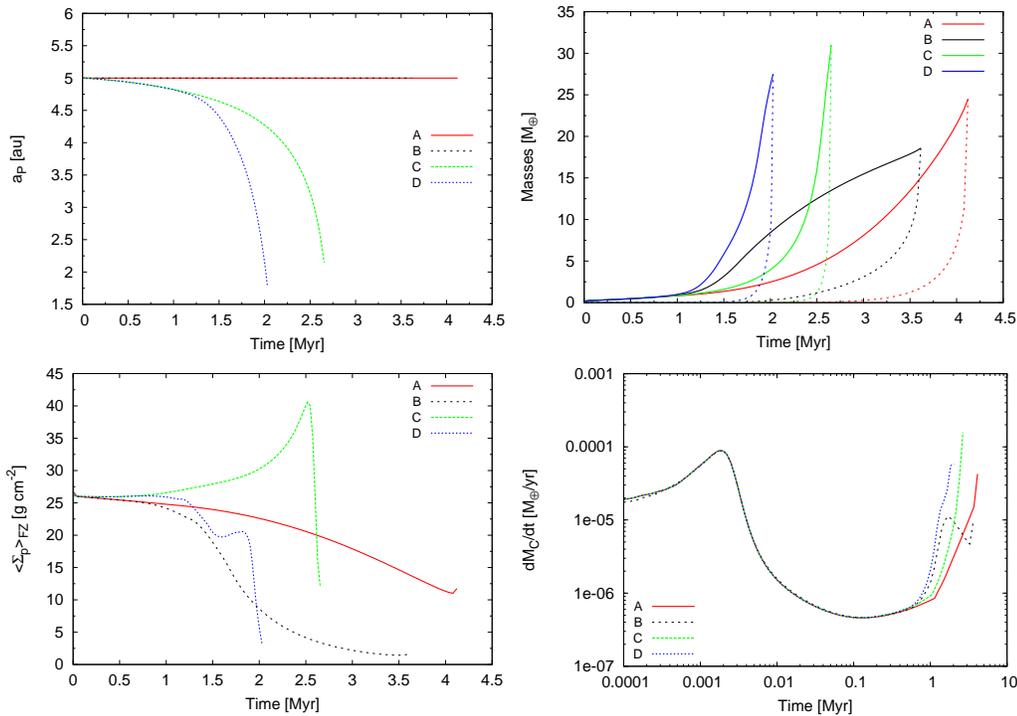} 
 \vspace*{-.6 cm}
 \caption{Time evolution of: planet's semi-major axis (top left panel), core mass and envelope mass (top right panel), total planetesimal surface density at the planet's feeding zone (bottom left panel), and total planetesimal accretion rate (bottom right panel). The solid line A corresponds to the case of in situ formation when planetesimal fragmentation is not considered (case a). The large dashed line B corresponds to the case of in situ formation when planetesimal fragmentation is considered (case b). The short dashed line C corresponds to the case when planet migration is considered ($f_I= 0.01$) but planetesimal fragmentation is not considered (case a), and the dotted line D corresponds to the case when planet migration ($f_I= 0.01$) and planetesimal fragmentation are considered (case b). Color figure only available in the electronic version.} 
 \label{fig2}
\end{center}
\end{figure}

In Fig.~\ref{fig1}, we plot (for the case a) the time evolution of the planet's semi-mayor axis (left panel) and the time evolution of core mass and envelope mass (right panel) for the case of in situ formation ($f_I= 0$), and for different values of the reduction factor of type I migration. We found that the planet quickly achieves the inner radius of the disk (at 0.7 au) if type I migration is not strongly reduced ($f_I= 0.01$) or not considered. Moreover, only when $f_I= 0$ and $f= 0.01$ the planet core reaches the critical mass before the dissipation of the disk (6 Myr). For these cases, when planet migration is considered in the model, the planet reaches the critical core mass in $\sim 2.65$~Myr, implying a reduction of $\sim 35$\% in time respect to the case of in situ formation. This is due to an increment in the planetesimal surface density at the planet's feeding zone as a consequence of the inward migration of the planet (Fig.~\ref{fig2}, bottom left panel, curve C). 

For case b, we considered only the cases when $f_I= 0.01$ and $f_I= 0$. Fig.~\ref{fig2} represents the time evolution of: the planet's semi-mayor axis (top left panel), core mass and envelope mass (top right panel), the mean value of the total planetesimal surface density at the planet feeding zone (bottom left panel), and the total planetesimal accretion rate (bottom right panel). When planetesimal fragmentation and planet migration are considered the planet reaches the critical core mass in $\sim 2$~Myr. This implies a reduction in the time of $\sim 25$\% in comparison to the case of planet migration without planetesimal fragmentation (curve C), $\sim 45$\% in comparison to the case of in situ formation considering planetesimal fragmentation (curve B), and $\sim 52$\% in comparison to the case of in situ formation without planetesimal fragmentation (curve A). We note that despite the total planetesimal surface density in the planet's feeding zone for the case when planetesimal fragmentation and planet migration are considered (bottom left panel, curve D) being smaller than the case when only planet migration is considered (bottom left panel, curve C), the time at which the planet reaches the critical core mass is shorter (top right panel, curves D and C, respectively). This is because the accretion of small fragments (when planetesimal fragmentation is considered) causes that the total planetesimal accretion rate becomes greater (bottom right panel, curves D and C, respectively). 

 \vspace*{-.4 cm}

\section{Conclusions}    
 
Our results are in concordance with those found by \cite[Alibert et al. (2005 b)]{Alibert.et.al.2005b}: the migration of the planet favors the formation of a massive core. The combination of planet migration and planetesimal fragmentation reduces the time at which the planet reaches the critical core mass more than 50\% in comparison to the case of in situ formation without planetesimal fragmentation. We remark that the accretion of small fragments (products of the planetesimal fragmentation) increases the total planetesimal accretion rate of the planet even if the planetesimal surface density is smaller than the case where planetesimal fragmentation is not considered. 

Finally, we note that if type I migration is not strongly reduced the planet quickly reaches the inner radius of the disk and does not reach the critical core mass. However, if moderate migration is considered, together with planetesimal fragmentation, the planet reaches the critical core mass in a few millon years. This result could have important implications linking models that invoke the need for an inward migration of a proto Jupiter (\cite[Walsh et al. 2011]{Walsh.et.al.2011}) and models that invoke the need for a population of initial big planetesimals (\cite[Morbidelli et al. 2009]{Morbidelli.et.al.2009}).     

\end{document}